\documentclass[journal]{IEEEtran}
\IEEEoverridecommandlockouts
\usepackage{setspace}
\usepackage{cite}
\usepackage{amsmath,amssymb,amsfonts,mathtools,bm}
\usepackage{amsthm}
\usepackage{algorithm}
\usepackage{algpseudocode}
\usepackage{graphicx}
\usepackage{textcomp}
\usepackage{xcolor,float}
\usepackage{tabularx}
\usepackage{textcomp}
\usepackage{diagbox}
\usepackage{svg}
\usepackage{booktabs}
\usepackage{subfigure}
\usepackage{hyperref}

\usepackage{graphicx}
\usepackage{makecell}
\usepackage{amsthm}
\usepackage{caption}
\usepackage{sidecap}
\usepackage{microtype}
\usepackage{booktabs} 
\usepackage{multirow}
\usepackage{float}
\usepackage{adjustbox} 
\usepackage{amsmath}
\usepackage{amssymb}
\usepackage{colortbl}
\usepackage{makecell}
\usepackage{float}
\usepackage{url}
\usepackage{balance}
\usepackage{bbding}
\usepackage{hyperref}
\usepackage{graphicx}
\usepackage{sidecap}
\usepackage{microtype}
\usepackage{graphicx}
\usepackage{subfigure}
\usepackage{booktabs} 
\usepackage{multirow}
\usepackage{array}
\usepackage{subcaption}
\usepackage{graphicx}
\usepackage{makecell}
\usepackage{amsmath}
\usepackage{amssymb}

\usepackage{tikz}

\usepackage{changes} 

\definecolor{lime}{HTML}{A6CE39}
\DeclareRobustCommand{\orcidicon}{%
    \begin{tikzpicture}
    \draw[lime, fill=lime] (0,0) 
    circle [radius=0.16] 
    node[white] {{\fontfamily{qag}\selectfont \tiny ID}};    \draw[white, fill=white] (-0.0625,0.095) 
    circle [radius=0.007];    \end{tikzpicture}
    \hspace{-2mm}}
\foreach \x in {A, ..., Z}{%
    \expandafter\xdef\csname orcid\x\endcsname{\noexpand\href{https://orcid.org/\csname orcidauthor\x\endcsname}{\noexpand\orcidicon}}
    }

\allowdisplaybreaks
\renewcommand{\baselinestretch}{1}

\begin{document}

\title{BeamAgent: LLM-Aided MIMO Beamforming with Decoupled Intent Parsing and Alternating Optimization for Joint Site Selection and Precoding}
\author{
Xiucheng Wang, Yue Zhang, Nan Cheng

\thanks{
\par This work was supported by the National Key Research and Development Program of China (2024YFB907500).
\par Xiucheng Wang, Yue Zhang, and Nan Cheng are with the State Key Laboratory of ISN and School of Telecommunications Engineering, Xidian University, Xi’an 710071, China (e-mail: \{xcwang\_1,25011210995\}@stu.xidian.edu.cn; dr.nan.cheng@ieee.org);\textit{(Corresponding author: Nan Cheng.)}. 

}

} 
    
    \maketitle

\IEEEdisplaynontitleabstractindextext

\IEEEpeerreviewmaketitle

\begin{abstract}
Integrating large language models (LLMs) into wireless communication optimization is a promising yet challenging direction. Existing approaches either use LLMs as black-box solvers or code generators, tightly coupling them with numerical computation. However, LLMs lack the precision required for physical-layer optimization, and the scarcity of wireless training data makes domain-specific fine-tuning impractical. We propose BeamAgent, an LLM-aided MIMO beamforming framework that explicitly decouples semantic intent parsing from numerical optimization. The LLM serves solely as a semantic translator that converts natural language descriptions into structured spatial constraints. A dedicated gradient-based optimizer then jointly solves the discrete base station site selection and continuous precoding design through an alternating optimization algorithm. A scene-aware prompt enables grounded spatial reasoning without fine-tuning, and a multi-round interaction mechanism with dual-layer intent classification ensures robust constraint verification. A penalty-based loss function enforces dark-zone power constraints while releasing optimization degrees of freedom for bright-zone gain maximization. Experiments on a ray-tracing-based urban MIMO scenario show that BeamAgent achieves a bright-zone power of 84.0\,dB, outperforming exhaustive zero-forcing by 7.1\,dB under the same dark-zone constraint. The end-to-end system reaches within 3.3\,dB of the expert upper bound, with the full optimization completing in under 2\,s on a laptop.
\end{abstract}




%

\section{Introduction}
\label{sec:introduction}

\IEEEPARstart{M}{ultiple-input} multiple-output (MIMO) beamforming is a key enabler for spectral efficiency and spatial multiplexing in fifth-generation (5G) and sixth-generation (6G) wireless systems \cite{larsson2014massive, goldsmith2005wireless}. In practical deployments, beamforming optimization involves two coupled decisions: discrete base station (BS) site selection and continuous precoding matrix design. These decisions rely heavily on accurate channel knowledge, which can be obtained from site-specific radio environment characterization such as channel knowledge maps \cite{zeng2024tutorial, zeng2021toward} or radio propagation models \cite{10640063, hoydis2023sionna}. As 6G networks move toward environment-aware communication \cite{dang2020should, shen2023toward, 6g}, the demand for efficient and accessible optimization tools continues to grow. Conventionally, engineers must manually translate deployment objectives into mathematical constraints, screen candidate BS locations, and tune algorithm parameters. This process demands deep domain expertise and considerable time. Recent advances in large language models (LLMs) have demonstrated strong capabilities in natural language understanding and structured reasoning \cite{zhao2023survey}. Several studies have explored the integration of LLMs with wireless network optimization \cite{10648594, hong2025comprehensive, zhou2024large_survey}, opening up a promising direction for lowering the barrier of access to complex beamforming design.

Several recent studies have explored this direction from different perspectives. ComAgent \cite{li2026comagent} employs a multi-LLM collaborative architecture in which literature, planning, coding, and scoring agents cooperatively generate solver-ready formulations and simulation code. WirelessAgent \cite{tong2025wirelessagent} builds a cognitive loop of perception, memory, planning, and action for autonomous network slicing management. WirelessAgent++ \cite{tong2026wirelessagent++} further automates agent workflow design using Monte Carlo tree search. The prompt-engineering-based real-time feedback and verification framework \cite{mehmood2026bridging} uses LLMs to directly output physical-layer optimization solutions and iteratively refines them through system feedback. CommLLM \cite{jiang2024large} coordinates multiple agents for knowledge retrieval, collaborative planning, and solution evaluation in communication system design. In addition, NetLLM \cite{wu2024netllm} adapts LLMs to networking tasks through fine-tuning, and in-context learning has been explored for power control \cite{zhou2024large_icl}. LLMs have also been applied to resource allocation \cite{noh2025adaptive, wu2025llm} and power control \cite{wang2026wireless}. These works demonstrate the potential of LLMs in wireless network optimization from various angles.

Despite their differences in architecture, these methods share a common design pattern: the LLM is deeply involved in the numerical computation process. Specifically, existing approaches assign the LLM one of three roles. The first is a code generator that produces optimization algorithms. The second is a black-box solver that directly outputs numerical solutions. The third is a decision engine that autonomously determines network configurations in a reasoning loop. This coupling introduces three problems. LLM outputs lack the numerical precision required for physical-layer optimization. Constraint satisfaction cannot be guaranteed due to the stochastic nature of LLM inference. System behavior becomes difficult to debug and reproduce. Furthermore, autonomous reasoning loops such as ReAct introduce unpredictability that is undesirable for beamforming tasks where numerical accuracy is critical.

A natural remedy might be to fine-tune LLMs on wireless-specific datasets so that they acquire domain knowledge for numerical tasks. However, labeled training data for physical-layer optimization are extremely scarce. Unlike natural language corpora, generating ground-truth pairs of channel conditions and optimal precoding vectors requires expensive ray-tracing simulations \cite{hoydis2023sionna} or field measurements. Existing efforts to build telecom-specific LLMs, such as TelecomGPT \cite{zou2025telecomgpt} and WirelessGPT \cite{yang2025wirelessgpt}, have demonstrated the difficulty of curating sufficient domain data. At the same time, general-purpose LLMs trained on internet text lack knowledge of channel models, antenna array responses, and propagation physics \cite{zhou2024large_survey}. Although recent work has attempted to adapt LLMs for path loss prediction \cite{feng2025physics} and multipath generation \cite{huang2025llm4mg}, these efforts require specialized fine-tuning and remain limited to narrow subtasks. Directly applying general-purpose LLMs to beamforming optimization leads to physically implausible outputs.

Our key insight is that LLMs excel at semantic understanding and spatial reasoning, while wireless domain knowledge is more naturally embedded in signal processing algorithms and physics-based channel models \cite{10764739, zeng2024tutorial,10764739,11278649,11083758}. Rather than forcing domain knowledge into the LLM through scarce data, we propose to decouple the two capabilities. Based on this insight, we propose BeamAgent, an LLM-aided MIMO beamforming framework that explicitly separates intent understanding from numerical optimization. The LLM serves solely as a semantic translator that converts natural language into structured constraints, a task well within its pre-trained capabilities. The optimization engine, which encodes wireless domain knowledge through the channel model and the objective function, independently handles numerical solving.

BeamAgent adopts a deterministic pipeline architecture consisting of four functional stages. In the first stage, an LLM parses the user's natural language input into a structured constraint set. A scene-aware prompt embeds the map geometry, receiver coordinates, and semantic-to-spatial mapping rules into the LLM context, enabling grounded spatial reasoning without domain-specific fine-tuning. In the second stage, a multi-round interaction mechanism allows the user to verify and refine the parsed constraints. An ASCII grid preview provides intuitive spatial visualization, and a dual-layer intent classifier combines LLM-based classification with keyword-rule fallback to ensure robustness when the LLM service is unavailable. In the third stage, an alternating optimization algorithm jointly solves the discrete BS site selection and continuous precoding design. A vectorized parallel search evaluates all candidate transmitter locations simultaneously, followed by gradient descent for precoding refinement. A penalty-based reformulation converts the dark-zone power constraint into a smooth surrogate that releases optimization degrees of freedom once the constraint is satisfied. In the fourth stage, the system generates visualization results and performance analysis. Throughout the entire pipeline, the LLM is invoked only in the first two stages. Its output passes through a deterministic post-processing chain before entering the optimizer. The optimization engine runs entirely independent of the LLM, ensuring that solution quality is unaffected by LLM inference variability. The main contributions of this paper are summarized as follows.
\begin{enumerate}
    \item We propose BeamAgent, the first LLM-aided MIMO beamforming framework that explicitly decouples semantic intent parsing from numerical optimization. The LLM acts as a semantic translator that converts natural language descriptions into structured constraints. A dedicated optimization engine independently performs numerical solving. This separation eliminates the impact of LLM numerical hallucinations on optimization results and ensures reproducible system behavior.

    \item We design a scene-aware prompt engineering scheme and a robust multi-round interaction mechanism. The prompt encodes the complete physical environment into the LLM context, enabling grounded spatial reasoning without fine-tuning. A dual-layer intent classifier with LLM-based classification and keyword-rule fallback ensures reliable constraint verification even under LLM service unavailability. An incremental constraint update mechanism allows the LLM to perform delta modifications on existing constraints, maintaining consistency across interaction rounds.

    \item We propose an alternating optimization algorithm for joint BS site selection and precoding design. The framework combines vectorized exhaustive search for discrete site selection with gradient descent for continuous precoding refinement. A penalty-based reformulation converts the dark-zone power constraint into a smooth surrogate loss. The conditional penalty term vanishes once the constraint is satisfied, releasing optimization degrees of freedom for bright-zone gain maximization.

    \item We validate BeamAgent in an urban scenario of 224\,m $\times$ 222\,m with a 4$\times$4 MIMO configuration. Experimental results show a bright-zone power of 84.0\,dB, outperforming exhaustive zero-forcing by 7.1\,dB under the same dark-zone constraint. The end-to-end system reaches within 3.3\,dB of the expert upper bound, with the full optimization completing in under 2\,s on a laptop.
\end{enumerate}
The remainder of this paper is organized as follows. Section~\ref{sec:related_work} reviews the related work. Section~\ref{sec:system_model} presents the system model and problem formulation. Section~\ref{sec:beamagent} describes the proposed BeamAgent framework. Section~\ref{sec:results} provides numerical results and analysis. Section~\ref{sec:conclusion} concludes the paper.

%

\section{Related Work}
\label{sec:related_work}

\subsection{LLM-Aided Wireless Network Optimization}
\label{sec:rw_llm_wireless}

The integration of LLMs into wireless communication has evolved rapidly since 2023 \cite{zhou2024large_survey, hong2025comprehensive}. Early work focused on in-context learning (ICL), where LLMs make optimization decisions through prompting alone. Zhou et al. \cite{zhou2024large_icl} demonstrated that formatting power control problems as natural language prompts with demonstration examples enables performance comparable to deep reinforcement learning. The LLM-xApp framework \cite{wu2025llm} applied optimization by prompting to O-RAN resource management, becoming the first LLM-powered xApp validated on a real 5G testbed. LLMs have also been explored for adaptive resource allocation \cite{noh2025adaptive} and power control \cite{wang2026wireless} using prompt-based approaches.

A parallel line of research has pursued domain-specific fine-tuning. TelecomGPT \cite{zou2025telecomgpt} established a pipeline for adapting general-purpose LLMs to telecom through continual pre-training on 3GPP documents and instruction tuning. WirelessGPT \cite{yang2025wirelessgpt} proposed a generative pre-trained multi-task framework for wireless communication. Recent work has also attempted to adapt LLMs for physical-layer tasks such as path loss prediction \cite{feng2025physics} and multipath generation \cite{huang2025llm4mg}. These fine-tuned methods achieve higher accuracy than prompting but require task-specific training data that is expensive to obtain in the wireless domain.

Multi-agent architectures have become the dominant paradigm for complex wireless tasks \cite{10648594}. WirelessAgent \cite{tong2025wirelessagent} implements a cognitive loop of perception, memory, planning, and action via LangGraph for network slicing management. ComAgent \cite{li2026comagent} coordinates five specialized agents in a perception-planning-action-reflection loop, achieving a 72\% solution success rate across 25 wireless tasks including MIMO beamforming. WirelessAgent++ \cite{tong2026wirelessagent++} further automates agent workflow design itself using Monte Carlo tree search over workflow topologies. CommLLM \cite{jiang2024large} deploys multi-agent retrieval, planning, and evaluation for communication system design. The PE-RTFV framework \cite{mehmood2026bridging} uses a dual-LLM architecture where one LLM generates structured prompts and another produces physical-layer solutions through iterative feedback.

Despite the diversity of these approaches, they share a common pattern: the LLM is deeply involved in the numerical computation process. In ICL and prompt-based methods, the LLM directly outputs numerical solutions whose precision is limited by tokenization. In fine-tuned models, the LLM replaces the conventional solver entirely. In multi-agent systems, the LLM either generates optimization code or autonomously makes decisions in a reasoning loop. BeamAgent departs from all three paradigms by restricting the LLM to semantic intent parsing and delegating all numerical computation to a dedicated optimization engine.

\subsection{Joint BS Site Selection and Precoding Optimization}
\label{sec:rw_joint}

Joint transmitter placement and precoding design is a classical problem in wireless network planning \cite{amaldi2003planning}. Conventional approaches decouple the two variables and solve them sequentially: the BS location is first determined through coverage analysis or combinatorial search, and the precoding matrix is then optimized for the fixed location. Alternating optimization has been widely used in related joint design problems, such as joint IRS phase shift and precoding \cite{wu2019intelligent} and joint antenna position and beamforming in movable antenna systems \cite{peng2024joint}. Radio environment characterization through channel knowledge maps \cite{zeng2024tutorial, zeng2021toward} and radio map construction \cite{10764739, levie2021radiounet, feng2025recent} provides the channel data foundation for such optimization. Ray-tracing simulators \cite{hoydis2023sionna} enable site-specific propagation prediction but are computationally expensive.

Data-driven methods have been applied to BS placement in recent years. Mallik and Villemaud \cite{mallik2025base} proposed a deep reinforcement learning framework for EMF-aware BS deployment using a conditional GAN as a digital twin. TelePlanNet \cite{deng2025teleplannet} applied LLMs with reinforcement learning for site selection, achieving 78\% planning-construction consistency. Qiu et al. \cite{qiu2024large} created an LLM-based wireless network design method integrated with ray-tracing propagation models, demonstrating the potential of LLMs in combinatorial site placement optimization.

However, existing joint optimization methods universally assume that the optimization constraints are specified by domain experts in advance. Which receivers should be enhanced, which should be suppressed, and where the transmitter may be placed are treated as given inputs. No prior work has addressed the front-end problem of extracting these constraints from natural language descriptions. BeamAgent fills this gap by integrating LLM-based intent parsing as a first-class component of the optimization pipeline.

\subsection{LLM-Based Intent Understanding for Network Optimization}
\label{sec:rw_intent}

Intent-driven networking has attracted increasing attention as a means to lower the barrier of network management. Mekrache et al. \cite{mekrache2024intent} proposed an LLM-centric approach for intent-based management of next-generation networks, demonstrating natural language to network service descriptor translation. The Confucius framework \cite{wang2025intent} presented multi-agent LLM intent-driven network management at ACM SIGCOMM 2025. Wang et al. \cite{wang2024netconfeval} proposed the NetConfEval framework that employs fine-tuned GPT models to translate network requirements into JSON and API calls for network configuration.

These intent-driven systems focus on network-layer tasks such as service orchestration, slice configuration, and protocol automation. The translation target is typically a network descriptor or an API call, which operates at a higher level of abstraction than physical-layer optimization parameters. In contrast, BeamAgent requires the LLM to perform spatial reasoning over a physical coordinate system. The translation target is a set of geometric constraints, including bounding box coordinates and maximize/minimize objectives, that directly parameterize a beamforming optimization problem. This grounded spatial reasoning task differs fundamentally from the network-layer intent parsing addressed by prior work and demands a scene-aware prompt design that encodes the deployment geometry into the LLM context.

\subsection{Summary and Positioning of This Work}
\label{sec:rw_summary}

Table~\ref{tab:comparison} summarizes the key differences between BeamAgent and representative existing methods. Three distinguishing features are highlighted.

First, BeamAgent explicitly decouples the LLM from numerical optimization. Existing methods assign the LLM a computational role as a code generator, a black-box solver, or an autonomous decision engine. BeamAgent restricts the LLM to semantic intent parsing. All numerical computation is performed by a dedicated gradient-based optimizer. This separation eliminates the impact of LLM numerical imprecision on solution quality.

Second, BeamAgent operates at the physical layer and jointly optimizes BS site selection and precoding design. Most LLM-aided wireless systems target network-layer tasks such as resource allocation, slicing, or protocol automation. Joint site selection and beamforming over site-specific ray-tracing channels has not been addressed by prior LLM-aided frameworks.

Third, BeamAgent introduces grounded spatial reasoning into the LLM pipeline. Existing intent-driven systems translate natural language into network descriptors or API calls. BeamAgent translates natural language into geometric constraints over a physical coordinate system. This requires the LLM to align semantic region descriptions with precise spatial coordinates, a capability enabled by our scene-aware prompt design without domain-specific fine-tuning.

\begin{table*}[t]
\centering
\caption{Comparison of BeamAgent with representative existing methods.}
\label{tab:comparison}
\resizebox{0.98\linewidth}{!}{
\begin{tabular}{@{}l|l|l|l|l|l@{}}
\toprule
\textbf{Method} & \textbf{LLM Role} & \textbf{Optimization Layer} & \textbf{NL Interface} & \textbf{Fine-Tuning} & \textbf{Solver} \\
\midrule
ComAgent \cite{li2026comagent} & Code generator & Task-dependent & No & No & LLM-generated code \\
WirelessAgent \cite{tong2025wirelessagent} & Decision engine & Network (slicing) & Partial & No & LLM reasoning loop \\
PE-RTFV \cite{mehmood2026bridging} & Black-box solver & Physical (constellation) & No & No & LLM output \\
NetLLM \cite{wu2024netllm} & End-to-end model & Network (scheduling) & No & Yes & Fine-tuned LLM \\
TelecomGPT \cite{zou2025telecomgpt} & Domain-adapted LLM & Telecom tasks & No & Yes & Fine-tuned LLM \\
TelePlanNet \cite{deng2025teleplannet} & Intent parser + RL & Site planning & Yes & Yes (RL) & RL policy \\
Qiu et al. \cite{qiu2024large} & Combinatorial optimizer & Site placement & No & No & LLM search \\
\midrule
\textbf{BeamAgent} & \textbf{Semantic translator} & \textbf{Physical (beamforming + site)} & \textbf{Yes} & \textbf{No} & \textbf{Dedicated gradient optimizer} \\
\bottomrule
\end{tabular}}
\end{table*}

\section{System Model and Problem Formulation}
\label{sec:system_model}
\begin{figure*}[ht]
    \centering
    \includegraphics[width=1\linewidth]{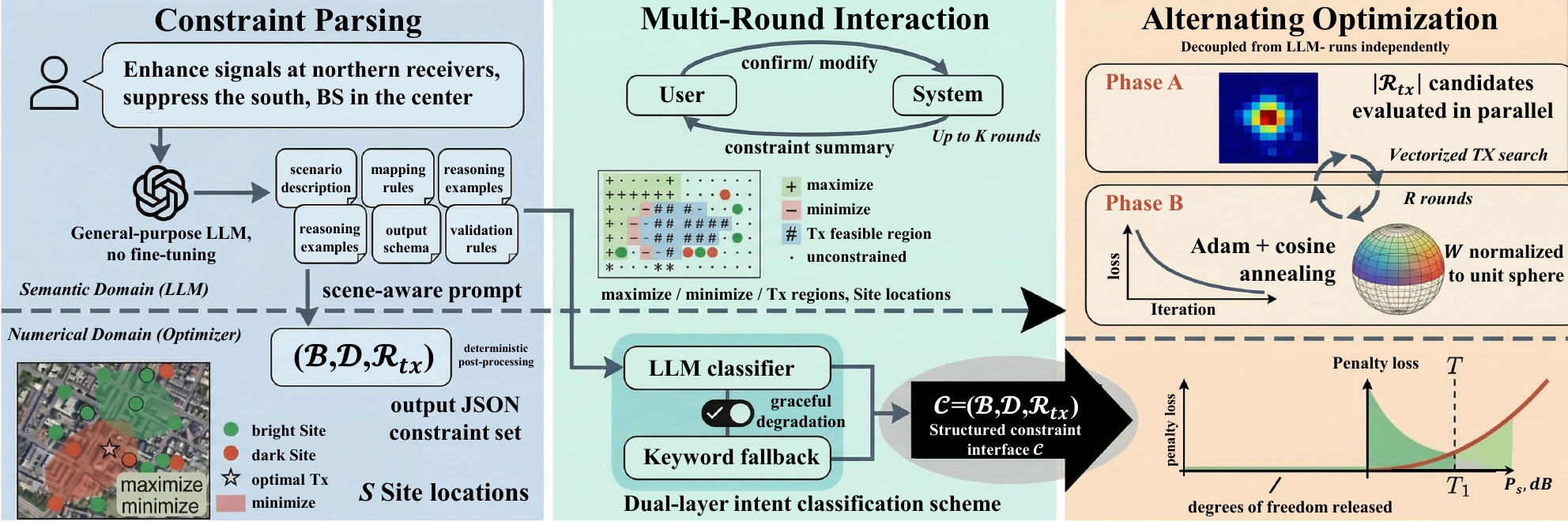}
    \caption{Overview of the BeamAgent framework. The system consists of three functional stages separated into two domains by the structured constraint interface $\mathcal{C} = (\mathcal{B}, \mathcal{D}, \mathcal{R}_{\mathrm{tx}})$. In the semantic domain (left), an LLM parses natural language input into spatial constraints via a scene-aware prompt, followed by multi-round interaction with dual-layer intent classification for constraint verification. In the numerical domain (right), an alternating optimization algorithm jointly solves BS site selection (Phase~A) and precoding design (Phase~B), with a threshold-based penalty that releases optimization degrees of freedom once the dark-zone constraint is satisfied.}
\label{fig:architecture}
\end{figure*}
\subsection{Physical Scenario and Ray-Tracing Data}
\label{sec:scenario}

We consider a urban area of size $L_x \times L_y$ operating at carrier frequency $f_c$. A total of $S$ receivers, referred to as Sites, are deployed at known fixed locations within the area. Each Site is indexed by $s \in \{1, \ldots, S\}$ with coordinates $(x_s, y_s)$. The transmitter is a single BS to be placed at one of $M$ candidate grid points. Each candidate location is indexed by $m \in \{1, \ldots, M\}$ with coordinates $(x_m, y_m)$. Both the transmitter and the receivers are equipped with $N$-element uniform linear arrays (ULAs) with half-wavelength spacing $d = \lambda / 2$.

Site-specific propagation data are obtained from ray-tracing simulations. For each Site $s$, the simulator provides multipath information to all $M$ candidate grid points. Each path $l$ is characterized by its field strength $E_l$ in dB$\mu$V/m, phase $\varphi_l$ in radians, departure angles $(\phi^{\mathrm{d}}_l, \theta^{\mathrm{d}}_l)$, and arrival angles $(\phi^{\mathrm{a}}_l, \theta^{\mathrm{a}}_l)$. Here $\phi$ denotes the azimuth angle and $\theta$ denotes the elevation angle. The number of resolvable paths per grid point varies depending on the local scattering environment.

\subsection{Ray-Tracing-Based MIMO Channel Model}
\label{sec:channel_model}

Based on the multipath data, we construct the MIMO channel matrix for each transmitter-receiver pair through coherent superposition. For a given candidate location $m$ and Site $s$, let $L_{m,s}$ denote the number of paths. The channel matrix $\mathbf{H}_{m,s} \in \mathbb{C}^{N \times N}$ is given by
\begin{equation}
\label{eq:channel}
\mathbf{H}_{m,s} = \sum_{l=1}^{L_{m,s}} \alpha_l \, \mathbf{a}_{\mathrm{rx}}(\phi^{\mathrm{a}}_l, \theta^{\mathrm{a}}_l) \, \mathbf{a}_{\mathrm{tx}}(\phi^{\mathrm{d}}_l, \theta^{\mathrm{d}}_l)^H,
\end{equation}
where $\alpha_l$ is the complex path gain and $\mathbf{a}_{\mathrm{rx}}$, $\mathbf{a}_{\mathrm{tx}} \in \mathbb{C}^{N}$ are the receive and transmit steering vectors, respectively.

The complex path gain encodes both the amplitude attenuation and the phase rotation of each path:
\begin{equation}
\label{eq:path_gain}
\alpha_l = 10^{E_l / 20} \cdot e^{j \varphi_l}.
\end{equation}

For an $N$-element ULA oriented along the reference direction $\theta_{\mathrm{orient}}$, the $n$-th element of the steering vector is
\begin{equation}
\label{eq:steering}
[\mathbf{a}(\phi, \theta)]_n = \exp\!\left( j \, 2\pi \, n \, \frac{d}{\lambda} \, u(\phi, \theta) \right), \quad n = 0, \ldots, N{-}1,
\end{equation}
where the spatial frequency is defined as $u(\phi, \theta) = \sin\theta \cdot \cos(\phi - \theta_{\mathrm{orient}})$.

In the ray-tracing simulation, each Site acts as the simulated transmitter and the grid points serve as simulated receivers. In our system, however, the grid points are candidate BS locations and the Sites are actual receivers. By electromagnetic reciprocity, the physical channel is obtained as
\begin{equation}
\label{eq:reciprocity}
\mathbf{H}_{m,s}^{\mathrm{phy}} = \mathbf{H}_{m,s}^T.
\end{equation}
All subsequent derivations use $\mathbf{H}_{m,s}^T$ as the effective channel. To enable gradient-based optimization, we decompose each channel matrix into its real and imaginary parts, i.e., $\mathbf{H} = \mathbf{H}_r + j\,\mathbf{H}_i$. All matrix operations are then expressed in real-valued arithmetic compatible with automatic differentiation frameworks.

\subsection{From User Intent to Optimization Constraints}
\label{sec:intent}

Conventional beamforming formulations assume that the optimization constraints are specified by domain experts in advance. In particular, which Sites should receive enhanced signals, which should be suppressed, and where the BS may be placed are treated as given inputs. In practice, however, such requirements often originate from non-expert users and are expressed in unstructured natural language. For example, a network planner may state ``enhance signals at northern receivers and suppress those in the south, with the BS placed in the center.'' To make the optimization system accessible to such users, a front-end mapping is required to convert natural language descriptions into structured optimization parameters.

We formalize this mapping as
\begin{equation}
\label{eq:parse}
f_{\mathrm{parse}}: \mathcal{T} \rightarrow \mathcal{C} = \left( \mathcal{B},\; \mathcal{D},\; \mathcal{R}_{\mathrm{tx}} \right),
\end{equation}
where $\mathcal{T}$ denotes the space of natural language inputs. The output constraint set $\mathcal{C}$ consists of three components: $\mathcal{B} \subseteq \{1, \ldots, S\}$ is the set of bright Sites whose received power should be maximized, $\mathcal{D} \subseteq \{1, \ldots, S\}$ is the set of dark Sites whose received power should be minimized, and $\mathcal{R}_{\mathrm{tx}} \subseteq \{1, \ldots, M\}$ is the set of candidate locations where the BS is allowed to be placed. 

This mapping involves two types of reasoning. The first is spatial grounding, which aligns semantic region descriptions such as ``northern'' or ``center'' with physical coordinate ranges. The second is intent classification, which interprets qualitative objectives such as ``enhance'' or ``suppress'' into formal maximize or minimize targets. Both tasks require an understanding of the physical deployment scenario and the ability to perform spatial inference. The existence of this front-end parsing problem provides the fundamental motivation for introducing an LLM into the beamforming optimization pipeline.

\subsection{Joint Optimization Problem Formulation}
\label{sec:problem}

Given the constraint set $\mathcal{C} = (\mathcal{B}, \mathcal{D}, \mathcal{R}_{\mathrm{tx}})$ obtained from the intent parsing stage, we formulate the joint BS site selection and precoding design problem. The optimization involves two coupled variables: the discrete BS location $m^* \in \mathcal{R}_{\mathrm{tx}}$ and the continuous precoding vector $\mathbf{W} \in \mathbb{C}^{N \times 1}$.

For a given BS location $m$ and Site $s$, the received power is
\begin{equation}
\label{eq:rx_power}
P_s(m, \mathbf{W}) = \left\| \mathbf{H}_{m,s}^T \mathbf{W} \right\|_F^2.
\end{equation}

The design objective is to maximize the average received power at bright Sites while keeping the power at all dark Sites below a prescribed threshold $T$. The joint optimization problem is formulated as
\begin{subequations}
\label{eq:joint_opt}
\begin{align}
\max_{m,\, \mathbf{W}} \quad & \frac{1}{|\mathcal{B}|} \sum_{s \in \mathcal{B}} P_s(m, \mathbf{W}) \label{eq:obj}\\
\mathrm{s.t.} \quad & P_{s,\mathrm{dB}}(m, \mathbf{W}) \leq T, \quad \forall\, s \in \mathcal{D}, \label{eq:dark_const}\\
& \|\mathbf{W}\| = 1, \label{eq:norm_const}\\
& m \in \mathcal{R}_{\mathrm{tx}}, \label{eq:site_const}
\end{align}
\end{subequations}
where $P_{s,\mathrm{dB}} = 10\log_{10} P_s$ is the received power in dB scale. Constraint \eqref{eq:dark_const} limits the interference at each dark Site. Constraint \eqref{eq:norm_const} enforces unit transmit power. Constraint \eqref{eq:site_const} restricts the BS placement to the feasible region specified by the user intent.

Problem \eqref{eq:joint_opt} is a mixed-integer nonconvex optimization. The discrete nature of $m$ precludes gradient-based methods over the site selection variable. The unit-norm constraint on $\mathbf{W}$ renders the continuous subproblem nonconvex even for a fixed BS location. Moreover, the two variables are coupled through the channel matrix $\mathbf{H}_{m,s}^T$, which depends on the selected location. In Section~\ref{sec:beamagent}, we present an alternating optimization framework that addresses these challenges. The dark-zone constraint \eqref{eq:dark_const} is handled through a penalty-based reformulation that converts the constrained problem into an unconstrained surrogate suitable for gradient descent.


\section{Proposed BeamAgent Framework}
\label{sec:beamagent}

\subsection{Overall Architecture}
\label{sec:architecture}

BeamAgent adopts a deterministic pipeline architecture that processes the user's natural language input and produces the optimized BS location and precoding vector. As illustrated in Fig.~\ref{fig:architecture}, the framework consists of four functional stages: constraint parsing, multi-round interaction, alternating optimization, and result analysis. The first stage invokes the LLM to translate the user's natural language description into a structured constraint set. The second stage allows the user to verify and refine the parsed constraints through visual previews and iterative feedback. The third stage solves the joint optimization problem \eqref{eq:joint_opt} using an alternating framework that handles the discrete and continuous variables separately. The fourth stage generates visualizations and computes theoretical performance bounds for benchmarking.

A central design principle of BeamAgent is the strict separation between the LLM and the optimization engine. The LLM is invoked only in the first two stages for semantic understanding tasks. Its output is a structured JSON object that passes through a deterministic post-processing chain before entering the optimizer. The optimization engine in the third stage operates entirely independently of the LLM. This decoupled design ensures that solution quality depends solely on the optimization algorithm, not on LLM inference variability.

\subsection{LLM-Based Constraint Parsing}
\label{sec:parsing}

The constraint parsing stage implements the mapping $f_{\mathrm{parse}}$ defined in \eqref{eq:parse}. We employ a general-purpose LLM without domain-specific fine-tuning. To enable grounded spatial reasoning, we design a scene-aware system prompt that encodes the complete physical environment into the LLM context.

The system prompt consists of five components. The first is a physical scenario description that specifies the map dimensions, the coordinate system definition, and the precise location of each Site. The second is a set of semantic-to-spatial mapping rules that associate natural language region descriptors with coordinate ranges. For instance, ``northwest'' is mapped to the bounding box covering the upper-left quadrant of the map. A total of ten standard region mappings are predefined to serve as reliable spatial anchors. The third component provides constraint reasoning examples that demonstrate how spatial descriptions are converted into formal constraints. The fourth specifies the output format as a strict JSON schema containing all required fields. The fifth defines validation rules including coordinate range limits and default value handling.

Given the user's natural language input, the LLM generates a JSON object containing three types of constraints. The receiver region constraint specifies a spatial region by its bounding box coordinates and an optimization objective of either maximize or minimize. The transmitter constraint defines the feasible region for BS placement. The receiver antenna constraint optionally specifies the antenna orientation of selected Sites. The LLM output is not used directly. Instead, it passes through a deterministic post-processing chain that performs the following operations in sequence: stripping Markdown formatting, parsing the JSON structure, clipping bounding box coordinates to the map boundary, normalizing objective keywords to canonical forms, filling default values for missing fields, and constructing a type-safe constraint object. This chain corrects minor formatting errors and coordinate deviations in the LLM output before it reaches the optimizer.

The predefined mapping rules do not restrict the LLM to a fixed set of regions. When the user's description does not match any predefined region, the LLM can freely reason about custom bounding box coordinates based on the scenario description. This hybrid design combines the reliability of rule-anchored parsing with the flexibility of free-form spatial reasoning.

\subsection{Multi-Round Interaction and Intent Classification}
\label{sec:interaction}

Natural language is inherently ambiguous. A single parsing pass may not fully capture the user's intent. To address this, we design a multi-round interaction mechanism that allows the user to verify and refine the parsed constraints before optimization begins.

After the initial parsing, the system presents a constraint summary together with an ASCII grid preview. The preview renders the constraint set as a character grid where different symbols represent maximize regions, minimize regions, Site locations, and the BS feasible area. The vertical axis is flipped to match the geographic coordinate convention. This visualization enables the user to verify the spatial correctness of the parsed constraints without inspecting the underlying JSON structure.

The user then provides feedback in natural language. The system classifies the feedback into one of two intents: confirm or modify. We employ a dual-layer intent classification strategy for robustness. The first layer invokes the LLM to perform semantic classification. The second layer serves as a keyword-rule fallback that activates when the LLM call fails. The fallback matches the user's input against a set of confirmation keywords and negation markers. A response is classified as confirm only if it matches a confirmation keyword and contains no negation marker. This graceful degradation ensures that the interaction loop remains functional even when the LLM service is temporarily unavailable.

If the intent is classified as modify, the system invokes the LLM to perform an incremental constraint update. Rather than re-parsing the entire input from scratch, the current constraint JSON and the user's modification request are jointly provided to the LLM. The prompt explicitly instructs the LLM to preserve unmodified constraints and change only the parts requested by the user. This delta-update mechanism maintains consistency across interaction rounds and avoids error accumulation from repeated full parsing. The interaction loop continues for up to $K$ rounds until the user confirms the constraints.

\subsection{Penalty-Based Problem Reformulation}
\label{sec:penalty}

To solve problem \eqref{eq:joint_opt} using gradient-based methods, we reformulate the constrained optimization into an unconstrained surrogate. The dark-zone constraint \eqref{eq:dark_const} is incorporated into the objective through a penalty term. The unit-norm constraint \eqref{eq:norm_const} is handled by normalizing $\mathbf{W}$ at each iteration.

The resulting loss function is
\begin{equation}
\label{eq:loss}
\mathcal{L} = -\frac{1}{|\mathcal{B}|}\sum_{s \in \mathcal{B}} \ln P_s + \frac{\lambda}{|\mathcal{D}|}\sum_{s \in \mathcal{D}} \left[\max\!\left(0,\; P_{s,\mathrm{dB}} - T\right)\right]^2,
\end{equation}
where $P_{s,\mathrm{dB}} = 10\log_{10} P_s$, $T$ is the dark-zone power threshold from \eqref{eq:dark_const}, and $\lambda$ is the penalty coefficient. The first term maximizes the average log-power at bright Sites. The logarithmic transformation converts the multiplicative power scale into an additive one, improving numerical stability across the wide dynamic range of received powers. The second term penalizes dark Sites only when their power exceeds $T$. Once all dark-zone powers satisfy the constraint, the penalty vanishes and the optimizer allocates all spatial degrees of freedom to bright-zone gain maximization. The quadratic form $[\max(0, \cdot)]^2$ provides smooth gradients near the threshold boundary.

\subsection{Alternating Optimization Algorithm}
\label{sec:ao}

Problem \eqref{eq:joint_opt} involves the coupled optimization of a discrete variable $m$ and a continuous variable $\mathbf{W}$. We adopt an alternating optimization strategy that iteratively solves two subproblems.

\subsubsection{Phase A: Vectorized BS Site Search}

Given a fixed precoding vector $\mathbf{W}$, the BS site selection reduces to evaluating the loss \eqref{eq:loss} at each candidate location and selecting the one with the minimum value:
\begin{equation}
\label{eq:phase_a}
m^* = \arg\min_{m \in \mathcal{R}_{\mathrm{tx}}} \; \mathcal{L}(m, \mathbf{W}).
\end{equation}
To avoid iterating over candidates sequentially, we pre-stack the channel matrices of all candidates into a four-dimensional tensor of size $|\mathcal{R}_{\mathrm{tx}}| \times |\mathcal{B}| \times N \times N$ for bright Sites, and similarly for dark Sites. The tensor is reshaped to merge the candidate and Site dimensions, and the received powers for all candidates are computed through a single batched matrix multiplication. This vectorized evaluation reduces the computational cost from $O(|\mathcal{R}_{\mathrm{tx}}|)$ sequential operations to a single tensor operation.

\subsubsection{Phase B: Gradient-Based Precoding Optimization}

Given a fixed BS location $m^*$, the precoding optimization becomes a continuous problem over $\mathbf{W}$. We parameterize $\mathbf{W}$ as an unconstrained complex vector with real and imaginary components $\mathbf{W}_r, \mathbf{W}_i \in \mathbb{R}^{N \times 1}$ stored as learnable parameters. At each forward pass, $\mathbf{W}$ is normalized to unit norm. The received power is computed using real-valued arithmetic:
\begin{align}
\label{eq:real_arith}
(\mathbf{H}\mathbf{W})_r &= \mathbf{H}_r \mathbf{W}_r - \mathbf{H}_i \mathbf{W}_i, \nonumber \\
(\mathbf{H}\mathbf{W})_i &= \mathbf{H}_r \mathbf{W}_i + \mathbf{H}_i \mathbf{W}_r, \\
P_s &= \|(\mathbf{H}\mathbf{W})_r\|^2 + \|(\mathbf{H}\mathbf{W})_i\|^2. \nonumber
\end{align}
The loss \eqref{eq:loss} is then minimized using the Adam optimizer with cosine annealing learning rate scheduling. Gradient clipping with a maximum norm of $g_{\max}$ is applied to prevent gradient explosion caused by the logarithmic term at small power values.

\subsubsection{Alternating Procedure}

The overall algorithm alternates between Phase A and Phase B for $R$ outer rounds with a total of $T_{\mathrm{step}}$ gradient steps. In each round, Phase A selects the best BS location under the current $\mathbf{W}$, and Phase B refines $\mathbf{W}$ for $T_{\mathrm{step}} / R$ gradient steps at the selected location. The Adam optimizer is shared across all rounds without resetting its momentum estimates, which promotes smoother convergence. After the final round of Phase B, a concluding Phase A scan is performed with the converged $\mathbf{W}$ to confirm that the BS location is consistent with the final precoding vector. The complete procedure is summarized in Algorithm~\ref{alg:ao}.

\begin{algorithm}[ht]
\caption{Alternating Optimization for Joint BS Site Selection and Precoding Design}
\label{alg:ao}
\begin{algorithmic}[1]
\Require Channel tensors $\{\mathbf{H}_{m,s}\}$, constraint set $\mathcal{C} = (\mathcal{B}, \mathcal{D}, \mathcal{R}_{\mathrm{tx}})$, outer rounds $R$, total gradient steps $T_{\mathrm{step}}$, learning rate $\eta$, penalty coefficient $\lambda$, threshold $T$, clipping norm $g_{\max}$
\Ensure Optimal BS location $m^*$, precoding vector $\mathbf{W}^*$
\State Initialize $\mathbf{W}_r, \mathbf{W}_i \sim \mathcal{N}(0, 1/N)$; normalize $\mathbf{W}$
\State Create Adam optimizer with learning rate $\eta$
\State Create cosine annealing scheduler with $T_{\max} = T_{\mathrm{step}}$
\For{$r = 1$ to $R$}
    \State \textbf{Phase A:} Compute $\mathcal{L}(m, \mathbf{W})$ for all $m \in \mathcal{R}_{\mathrm{tx}}$ via batched tensor operations
    \State $m^* \leftarrow \arg\min_{m} \mathcal{L}(m, \mathbf{W})$
    \State Extract $\mathbf{H}_{m^*,s}$ for all $s \in \mathcal{B} \cup \mathcal{D}$
    \For{$t = 1$ to $T_{\mathrm{step}} / R$}
        \State \textbf{Phase B:} Normalize 
        $\mathbf{W} \leftarrow \mathbf{W} / \|\mathbf{W}\|$
        \State Compute $P_s$ for all $s \in \mathcal{B} \cup \mathcal{D}$ via \eqref{eq:real_arith}
        \State Compute $\mathcal{L}$ via \eqref{eq:loss}
        \State Backpropagate; clip gradients by $g_{\max}$
        \State Adam update; scheduler step
    \EndFor
\EndFor
\State \textbf{Final scan:} $m^* \leftarrow \arg\min_{m} \mathcal{L}(m, \mathbf{W})$
\State \Return $m^*$, $\mathbf{W} / \|\mathbf{W}\|$
\end{algorithmic}
\end{algorithm}


\section{Enperiment Results}
\label{sec:results}

\subsection{Experimental Setup}
\label{sec:setup}

We evaluate BeamAgent on an urban scenario with map dimensions $L_x = 224$\,m and $L_y = 222$\,m at carrier frequency $f_c = 2$\,GHz. The system deploys $S = 4$ fixed receiver Sites and considers $M = 33{,}089$ candidate grid points for BS placement. Both the transmitter and receivers are equipped with $N = 4$-element ULAs with half-wavelength spacing. Site-specific multipath propagation data are obtained from WinProp ray-tracing simulations. Each Site generates a binary file of 9--23\,MB containing 4--30 paths per grid point. Channel matrices are constructed via the coherent superposition model in \eqref{eq:channel} and cached in compressed NumPy format. The first computation takes approximately 9\,s; subsequent loads from cache complete in 0.3\,s. The default optimization hyperparameters are: $R = 5$ alternating rounds, $T_{\mathrm{step}} = 400$ total gradient steps, learning rate $\eta = 0.05$, penalty coefficient $\lambda = 0.1$, dark-zone threshold $T = 30$\,dB, and gradient clipping norm $g_{\max} = 5.0$. The LLM used for constraint parsing and intent classification is Claude Sonnet 4.6, accessed through the OpenAI-compatible API with max tokens of 1024 for parsing and 64 for classification. All optimization experiments are conducted on an Apple M2 MacBook.

We compare BeamAgent against five baselines. Random precoding samples a unit-norm vector from $\mathcal{CN}(0, \mathbf{I})$. Exhaustive MRT (Exh+MRT) selects the BS location that maximizes bright-zone power using the dominant right singular vector. Exhaustive MMSE (Exh+MMSE) applies minimum mean squared error precoding at each candidate location. Exhaustive ZF (Exh+ZF) projects the MRT solution onto the null space of the dark-zone channel. Signal-to-leakage-plus-noise ratio (SLNR) maximizes the ratio of bright-zone power to dark-zone leakage. To ensure fair comparison, all baselines are post-processed by scaling $\|\mathbf{W}\|$ so that $\max_{s \in \mathcal{D}} P_{s,\mathrm{dB}} \leq T$. For each Site $s$, the SVD of $\mathbf{H}_s$ yields the per-site achievable range $[\sigma_{\min,s}^2, \sigma_{\max,s}^2]$, and the theoretical contrast upper bound is $C_{\mathrm{UB}} = \min_{s \in \mathcal{B}} \sigma_{\max,s}^2 - \max_{s \in \mathcal{D}} \sigma_{\min,s}^2$. The default test constraint assigns Sites~7 and~8 as bright, Sites~6 and~12 as dark, with the BS restricted to the center region.

\subsection{Optimization Performance}
\label{sec:opt_perf}

Table~\ref{tab:method_comparison} compares all methods under the constraint $T = 30$\,dB. BeamAgent achieves a mean bright-zone power of 84.0\,dB, which is 5.4\,dB higher than the second-best Exh+MMSE (78.6\,dB) and 7.1\,dB higher than Exh+ZF (76.9\,dB). The scale column reveals that the original BeamAgent output requires only 0.9\,dB of post-processing reduction, meaning the optimizer has nearly satisfied the dark-zone constraint on its own. In contrast, Random and Exh+MRT require 19.9 and 20.3\,dB of scaling, indicating that these methods do not account for dark-zone suppression. Exh+ZF and SLNR suppress the dark zone to 14.5 and 14.3\,dB, far below the threshold, thereby wasting spatial degrees of freedom that could have been used to increase bright-zone gain.

\begin{table}[t]
\centering
\caption{Comparison of constraint-aware precoding methods in a 4-Site scenario with dark-zone threshold $T=30$\,dB. Methods marked with $^{\dagger}$ are post-processed by scaling $\|\mathbf{W}\|$ to enforce $\max_{s\in\mathcal{D}} P_{s,\text{dB}} \leq T$. The Scale column indicates the dB reduction applied during post-processing.}
\label{tab:method_comparison}
\resizebox{0.98\linewidth}{!}{
\begin{tabular}{@{}l|c|c|c|c@{}}
\toprule
Method & $\bar{P}_{\mathcal{B}}$ (dB) & $\max P_{\mathcal{D}}$ (dB) & Scale (dB) & Contrast (dB) \\
\midrule
Random & 75.9 & 30.0\,$^{\dagger}$ & 19.9 & 43.2 \\
Exh+MRT & 77.8 & 30.0\,$^{\dagger}$ & 20.3 & 48.8 \\
Exh+MMSE & 78.6 & 30.0\,$^{\dagger}$ & 1.0 & 57.7 \\
Exh+ZF & 76.9 & 14.5 & --- & 70.1 \\
SLNR & 76.4 & 14.3 & --- & 80.9 \\
\textbf{BeamAgent} & \textbf{84.0} & 30.0\,$^{\dagger}$ & \textbf{0.9} & \textbf{54.2} \\
\bottomrule
\end{tabular}}
\end{table}

\begin{figure}[t]
\centering
\includegraphics[width=\linewidth]{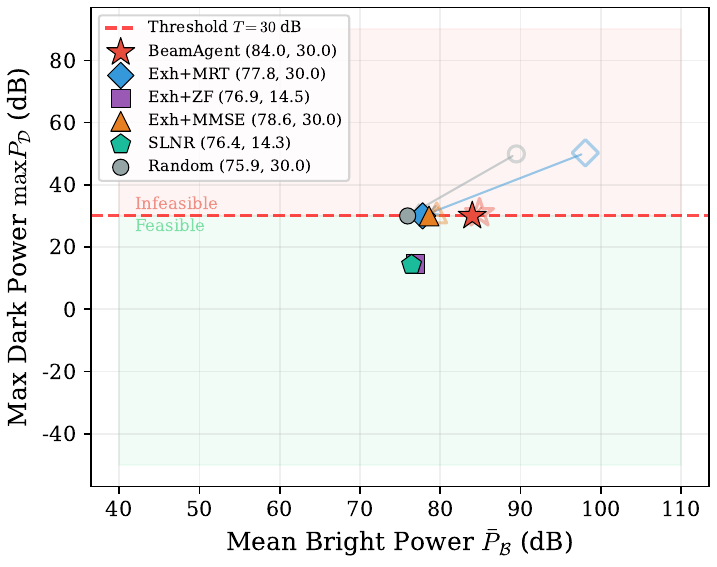}
\caption{Bright-zone power $\bar{P}_{\mathcal{B}}$ versus maximum dark-zone power $\max P_{\mathcal{D}}$ for all methods after post-processing. The dashed line marks the dark-zone threshold $T=30$\,dB. Points below this line are feasible. BeamAgent achieves the highest $\bar{P}_{\mathcal{B}}$ in the feasible region while operating near the constraint boundary.}
\label{fig:pareto}
\end{figure}

Fig.~\ref{fig:pareto} visualizes this tradeoff in a two-dimensional space with $\bar{P}_{\mathcal{B}}$ on the horizontal axis and $\max P_{\mathcal{D}}$ on the vertical axis. The horizontal dashed line at $T = 30$\,dB separates the feasible region (below) from the infeasible region (above). BeamAgent occupies the rightmost position in the feasible region, confirming that it achieves the highest bright-zone power among all methods that satisfy the dark-zone constraint. Exh+ZF and SLNR lie deep in the feasible region with low dark-zone power but also lower bright-zone power. This illustrates the key advantage of the threshold-based penalty: it drives the optimizer to the constraint boundary rather than over-suppressing the dark zone.

\begin{figure}[t]
\centering
\includegraphics[width=\linewidth]{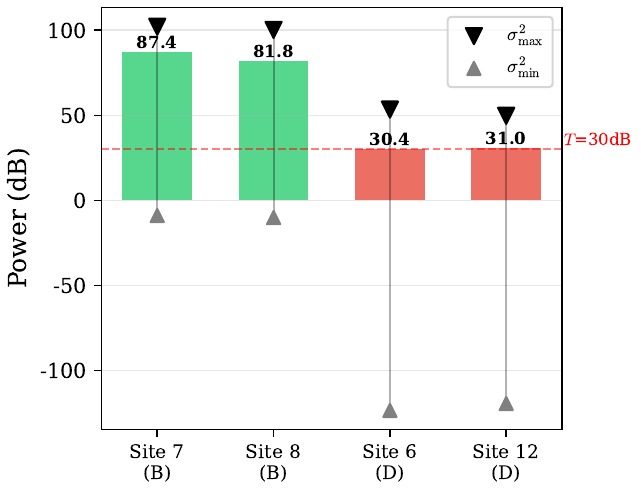}
\caption{Per-site received power (bars) and SVD bounds (triangles) at the optimized BS location. Black triangles ($\blacktriangledown$) denote $\sigma_{\max}^2$ and gray triangles ($\vartriangle$) denote $\sigma_{\min}^2$. The dashed line marks $T=30$\,dB. Bright Sites approach their upper bounds; dark Sites are suppressed to the threshold.}
\label{fig:persite}
\end{figure}

Fig.~\ref{fig:persite} shows the per-site received power at the optimized BS location alongside the SVD theoretical bounds. The bright Sites (7 and 8) achieve 87.4 and 81.8\,dB, both close to their respective $\sigma_{\max}^2$ values. The dark Sites (6 and 12) are suppressed to 30.4 and 31.0\,dB, tightly tracking the threshold $T = 30$\,dB. The large gap between $\sigma_{\max}^2$ and $\sigma_{\min}^2$ at dark Sites indicates that the channel provides sufficient dynamic range for selective suppression.

\begin{figure}[t]
\centering
\includegraphics[width=\linewidth]{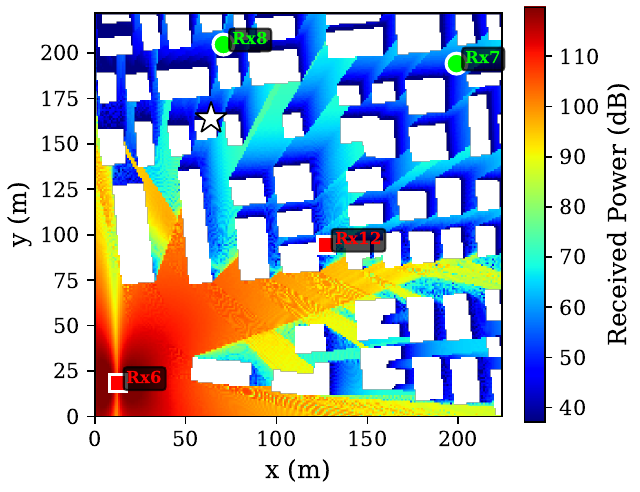}
\caption{Spatial distribution of received power (dB) in the 224\,m $\times$ 222\,m urban scenario after joint optimization. Green circles indicate bright Sites (7, 8); red squares indicate dark Sites (6, 12); the star marks the optimized BS location. The beam pattern steers energy toward the north while suppressing the south and center.}
\label{fig:heatmap}
\end{figure}

Fig.~\ref{fig:heatmap} presents the spatial power distribution after optimization. The beam pattern concentrates energy toward the bright Sites in the northern region while maintaining low power toward the dark Sites in the south and center. The optimized BS location is marked by a star symbol in the upper-left area.

\subsection{End-to-End System Evaluation}
\label{sec:e2e}

To assess the impact of LLM parsing on overall system performance, we compare four integration modes in Table~\ref{tab:e2e}. Expert+Optimizer uses manually specified constraints with our alternating optimization algorithm, serving as the performance upper bound. BeamAgent replaces the expert with LLM parsing. LLM Direct~W asks the LLM to output the precoding vector directly without channel information. LLM Gen Code asks the LLM to generate optimization code, which produces an SLNR-based solver.

\begin{table}[t]
\centering
\caption{Comparison of four LLM integration modes using Claude Sonnet 4.6. Expert+Optimizer serves as the upper bound with manually specified constraints. BeamAgent uses LLM-parsed constraints with the same optimizer. LLM Direct~W and LLM Gen Code represent alternative paradigms where the LLM directly participates in numerical computation.}
\label{tab:e2e}
\resizebox{0.98\linewidth}{!}{
\begin{tabular}{@{}l|c|c|l@{}}
\toprule
Method & Contrast (dB) & Std (dB) & Note \\
\midrule
Expert+Optimizer & 55.3 & 0.0 & Expert constraints \\
\textbf{BeamAgent} & \textbf{52.0} & 3.6 & LLM parsing + optimizer \\
LLM Direct W & 49.4 & 0.0 & LLM outputs $\mathbf{W}$ directly \\
LLM Gen Code & 80.9\,$^{\ddagger}$ & 0.0 & LLM generates SLNR solver \\
\bottomrule
\multicolumn{4}{l}{\footnotesize $^{\ddagger}$$\bar{P}_{\mathcal{B}}=76.4$\,dB, lower than BeamAgent (84.0\,dB); no site selection.}
\end{tabular}}
\end{table}

\begin{figure}[t]
\centering
\includegraphics[width=\linewidth]{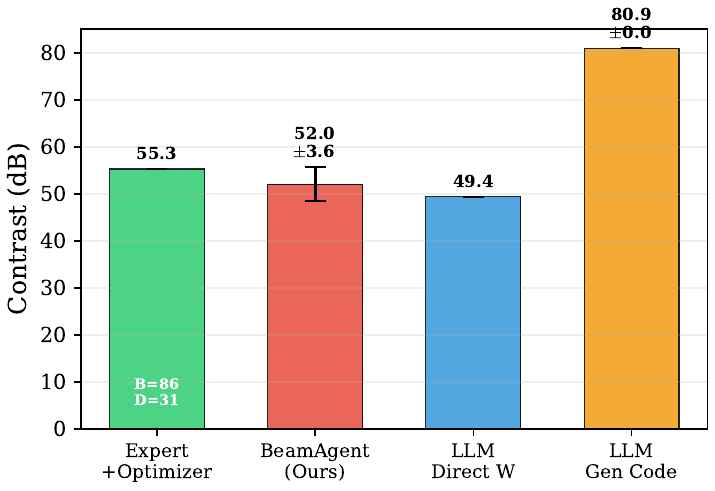}
\caption{Bright-dark contrast (dB) under four LLM integration modes using Claude Sonnet 4.6. BeamAgent (decoupled parsing + dedicated optimizer) achieves within 3.3\,dB of the expert upper bound. LLM Direct~W, which outputs the precoding vector without channel information, yields the lowest contrast among LLM-involved modes.}
\label{fig:e2e}
\end{figure}

BeamAgent achieves 52.0\,dB, within 3.3\,dB of the expert upper bound (55.3\,dB). The gap is caused by LLM parsing errors in identifying the correct bright/dark Site assignments. Over 10 independent runs, BeamAgent produces a mean contrast of 53.6\,dB with a standard deviation of 2.75\,dB, indicating reasonable reproducibility. The variance originates from the stochastic nature of LLM inference rather than the optimization algorithm.

LLM Direct~W (49.4\,dB) performs worst among the three LLM-involved modes. Without access to channel matrices, the LLM cannot determine meaningful precoding coefficients. This confirms that LLMs are not suitable for direct numerical decision-making in physical-layer optimization. LLM Gen Code produces an SLNR solver that achieves 80.9\,dB contrast. However, this high contrast comes at the cost of over-suppressing the dark zone to $-4.5$\,dB, far below the threshold, and achieving a bright-zone power of only 76.4\,dB. It also does not perform joint site selection. These results validate the decoupled design of BeamAgent: the LLM handles intent parsing where it excels, and the dedicated optimizer handles numerical solving where precision matters.

\subsection{LLM Constraint Parsing Evaluation}
\label{sec:parsing_eval}

We evaluate the constraint parsing accuracy of Claude Sonnet 4.6 on 25 test prompts spanning three difficulty levels: easy (10 prompts with standard region descriptions), medium (8 prompts with compound spatial references), and hard (7 prompts with implicit constraints and ambiguous language). Accuracy is measured by Bright IoU and Dark IoU between the parsed and ground-truth Site sets, and by end-to-end (E2E) success rate defined as both IoU values exceeding 0.5.

The overall E2E success rate is 64\%, with Bright IoU of 0.76 and Dark IoU of 0.71. Performance degrades with difficulty: easy prompts achieve 70\% E2E, medium 62\%, and hard 57\%. The most common failure mode is incorrect spatial grounding, where the LLM maps a region description to a bounding box that excludes the intended Site.

We conduct a prompt ablation study to assess the contribution of each prompt component. Removing the scene description reduces E2E from 60\% to 40\%, making it the most critical component. Removing reasoning examples reduces E2E to 50\%. The minimal prompt, which retains only the basic instruction, achieves an E2E of 60\% but with IoU values of 0.48 and 0.47, meaning the parsed constraints are spatially inaccurate even when the Site assignment is correct.

These results show that single-pass parsing is insufficient for reliable operation. The multi-round interaction mechanism described in Section~\ref{sec:interaction} addresses this limitation. Approximately 36\% of inputs require user correction through the ASCII grid preview. The dual-layer intent classifier ensures that the interaction loop remains functional even when the LLM service is temporarily unavailable.

\subsection{Parameter Sensitivity}
\label{sec:sensitivity}

\begin{figure}[t]
\centering
\includegraphics[width=\linewidth]{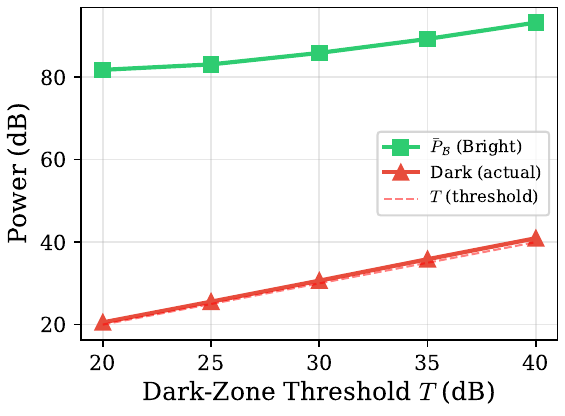}
\caption{Effect of dark-zone threshold $T$ on bright-zone power $\bar{P}_{\mathcal{B}}$ (green squares) and actual dark-zone power $\max P_{\mathcal{D}}$ (red triangles). The dashed line shows the threshold value. As $T$ increases, the dark-zone power tracks the threshold and bright-zone power rises accordingly, confirming that relaxing the constraint releases optimization degrees of freedom.}
\label{fig:threshold}
\end{figure}

Fig.~\ref{fig:threshold} shows the effect of the dark-zone threshold $T$ on optimization performance. As $T$ increases from 20 to 40\,dB, the actual dark-zone power closely tracks the threshold line, confirming that the penalty term effectively enforces the constraint. Simultaneously, the bright-zone power increases from 81 to 93\,dB. This behavior validates the core design principle of the threshold-based loss: relaxing the dark-zone constraint releases spatial degrees of freedom that the optimizer redirects toward bright-zone gain maximization.

The penalty coefficient $\lambda$ controls the tradeoff between bright-zone maximization and dark-zone constraint enforcement. Over the range $\lambda \in [0.01, 1.0]$, bright-zone power varies by less than 6\,dB. For $\lambda \geq 0.1$, the dark-zone power converges to the threshold. A small $\lambda = 0.01$ yields a dark-zone power of approximately 39\,dB, violating the constraint. The default value $\lambda = 0.1$ achieves a favorable balance.

The number of alternating rounds $R$ has minimal impact on performance for $R \geq 1$. Over the range $R \in \{1, 2, 3, 5, 10\}$, bright-zone power remains stable at approximately 86\,dB and dark-zone power stays near the threshold. The insensitivity to $R$ suggests that a single round of Tx--$\mathbf{W}$ alternation is sufficient for this problem scale, though $R = 5$ provides an additional consistency check.

\subsection{Ablation Study and Computational Efficiency}
\label{sec:ablation}

Table~\ref{tab:ablation} reports the ablation results. Removing the alternating optimization ($R=1$) causes the largest performance drop: bright-zone power decreases from 85.8 to 77.6\,dB ($-8.2$\,dB) and contrast drops from 55.1 to 46.7\,dB ($-8.4$\,dB). This confirms that joint optimization of BS location and precoding is essential. In contrast, removing cosine annealing or gradient clipping produces changes below 0.2\,dB, indicating that the Adam optimizer is sufficiently robust for this problem scale. The tradeoff loss variant (no threshold penalty) achieves a slightly higher bright-zone power (86.0\,dB) but suppresses the dark zone to 25.2\,dB, which is 4.8\,dB below the threshold. This over-suppression wastes degrees of freedom compared to the threshold mode, which precisely targets the constraint boundary.

\begin{table}[t]
\centering
\caption{Ablation study on optimization components in the 4-Site scenario ($T=30$\,dB). Each row removes one component from the full configuration. Disabling alternating optimization ($R=1$) causes the only significant performance degradation ($-8.2$\,dB in $\bar{P}_{\mathcal{B}}$).}
\label{tab:ablation}
\resizebox{0.98\linewidth}{!}{
\begin{tabular}{@{}l|c|c|c@{}}
\toprule
Configuration & $\bar{P}_{\mathcal{B}}$ (dB) & $\max P_{\mathcal{D}}$ (dB) & Contrast (dB) \\
\midrule
Full (Ours) & \textbf{85.8} & 30.6 & 55.1 \\
No Cosine Annealing & 85.9 & 30.7 & 55.2 \\
No Gradient Clipping & 85.8 & 30.6 & 55.2 \\
No Cosine + No Clip & 85.9 & 30.7 & 55.3 \\
$R=1$ (No Alternating) & 77.6 & 30.9 & 46.7 \\
Tradeoff (no penalty) & 86.0 & 25.2 & 60.7 \\
\bottomrule
\end{tabular}}
\end{table}

Table~\ref{tab:timing} reports the wall-clock time on an Apple M2 MacBook. The full optimization completes in 1.67\,s, of which constraint preparation (channel tensor stacking) accounts for 1.35\,s and gradient-based precoding optimization accounts for 0.31\,s. Each Phase~A Tx scan takes only 6\,ms, confirming the efficiency of the vectorized evaluation. For reference, exhaustive MRT over all candidate locations requires 0.12\,s. LLM API call latency is excluded as it depends on the external service and is not part of the local optimization pipeline.

\begin{table}[t]
\centering
\caption{Wall-clock time for each optimization stage measured on an Apple M2 MacBook. LLM API latency is excluded. The full pipeline completes in under 2\,s, with vectorized Tx scanning taking 6\,ms per round.}
\label{tab:timing}
\resizebox{0.98\linewidth}{!}{
\begin{tabular}{@{}l|c@{}}
\toprule
Stage & Time (s) \\
\midrule
Constraint preparation (tensor stacking) & 1.350 \\
Phase A: Tx scan (per round) & 0.006 \\
Phase B: 400 gradient steps & 0.308 \\
Full optimization (5 rounds) & 1.668 \\
Exhaustive MRT baseline (all Tx) & 0.124 \\
\bottomrule
\end{tabular}}
\end{table}

\section{Conclusion}
\label{sec:conclusion}
 
In this paper, we have proposed BeamAgent, an LLM-aided MIMO beamforming framework that decouples semantic intent parsing from numerical optimization. We have demonstrated that a general-purpose LLM, guided by a scene-aware prompt and multi-round interaction, can translate natural language into structured spatial constraints that drive a gradient-based alternating optimizer for joint BS site selection and precoding design. This decoupled architecture enables non-expert users to access physical-layer beamforming optimization through natural language, lowering the expertise barrier for wireless network planning and deployment. Future work will extend BeamAgent to multi-stream precoding, multi-BS coordination, dynamic channel scenarios, and on-device lightweight LLMs for reduced API dependency.

\bibliography{ref}
\bibliographystyle{IEEEtran}
\ifCLASSOPTIONcaptionsoff
  \newpage
\fi
\end{document}